\documentclass[showpacs,amsmath,amssymb,nofootinbib,prd]{revtex4}
\usepackage{graphicx}
\usepackage{enumerate}
\usepackage{amsmath}
\usepackage{amssymb}
\usepackage{amsfonts}
\usepackage{color}
\usepackage{pstricks}
\usepackage{color}
\newcommand{\Lagr}{\mathcal{L}}

\begin{document}
\title{Top quark Chromoelectric and Chromomagnetic Dipole Moments in a
Two Higgs Doublet Model with CP violation.}

\author{R. Gait\'an}
\email{rgaitan@unam.mx} \affiliation{Departamento de F\'isica, FES-Cuautitl\'an, UNAM, C.P. 54770, Estado de M\'exico, M\'exico}
\author{E. A. Garc\'es}
\email{egarces@fis.cinvestav.mx}\affiliation{Departamento de F\'isica, FES-Cuautitl\'an, UNAM, C.P. 54770, Estado de M\'exico, M\'exico}
\author{R. Martinez}
\email{remartinezm@unal.edu.co} \affiliation{Departamento de F\'isica, Universidad Nacional de Colombia, Bogot\'a D.C., Colombia}
\author{J.H. Montes de Oca}
\email{josehalim@gmail.com}\affiliation{Departamento de F\'isica, FES-Cuautitl\'an, UNAM, C.P. 54770, Estado de M\'exico, M\'exico}

\begin{abstract}
In this work we study the anomalous top quark-gluon couplings Chromoelectric Dipole Moment (CEDM) and Chromomagnetic Dipole Moment (CMDM)  in
a general THDM with CP violation. We find that this model provides an important contribution from the $Y_{tt}$ Yukawa coupling that needs to be
taken into account. The prediction for CMDM and CEDM obtained are $-0.03 < \Delta\tilde{k}_t < -0.005$ and $|\Delta \tilde{d}_t| < 0.005$, respectively.
\end{abstract}

\maketitle
\section{Introduction}
\label{sec-intro}

The recent discovery of a Higgs Boson with a mass of 125 GeV in the Large Hadron Collider (LHC)~\cite{Aad:2012tfa,Chatrchyan:2012ufa} has
confirmed that the Standard Model (SM) is the theory that correctly describes electroweak interactions. However, in the context of the SM there are many unsolved problems,
 among them one can mention:  (a) the fermion mass hierarchy, where the top quark is much heavier than other fermions, being the heaviest particle with a mass around the
symmetry breaking scale, (b) CP symmetry breaking in the CKM matrix cannot explain the matter antimatter problem and the
fermionic electric moments.This issues make very important the study of new sources of CP violation beyond the SM.

On the other hand, top quark physics is a relevant scenario for the study of physics beyond the SM~\cite{Ayazi:2013cba}.
In the next LHC run, at $\sqrt{s}=14$~TeV,  millions of top pairs will be produced, giving a great opportunity
for the study of top quark properties , including its couplings to the gauge fields and providing an excellent scenario
for new physics searches.

In the SM, top quark  magnetic and chromomagnetic dipole moments  (MDM and CMDM) are induced at one loop level; on the other hand, top quark anomalous electric and
chromoelectric  dipole moments (EDM and CEDM) appear only at 3 loop level, arising from the complex phase in the CKM matrix. This couplings are important in the study of
new physics; in order to generate them, extended models with new sources of CP violation are required. This anomalous couplings sensitively affect top pair production in $pp$ collisions.
Indirect bounds to the top quark MDM have been found from the bottom quark radiative decay $b\to s\gamma$~\cite{Hewett:1993em,Martinez:1996cy}, and also semileptonic B meson decays $(B\to Kl^{+}l^{-})$ have been used to improve indirect bounds to MDM and EDM~\cite{Kamenik:2011dk}.

Since anomalous top quark couplings $(t{t}g)$ affect the top  production, they have been widely studied at hadron colliders. Anomalous couplings have been studied in
top pair production \cite{Gupta:2009eq,Gupta:2009wu,Atwood:1991ka,Atwood:1994vm,Hioki:2009hm,Hioki:2013hva,Franzosi:2015osa,Biswal:2012dr}, top pair plus jets~\cite{Cheung:1995nt}, direct photon production~\cite{Hesari:2014hva} and single top production~\cite{Rizzo:1995uv}.
Also, spin correlations in top pair production have been used in the study
of CMDM\cite{Cheung:1996kc}. Constraints to CMDM and CEDM from Higgs boson production at the LCH have also been reported~\cite{Hayreter:2013kba}.

Anomalous moments have been calculated in different new physics scenarios
as in the case of MSSM~\cite{Yang:1996dma}, THDM~\cite{Martinez:2001qs}, Little Higgs model~\cite{Cao:2008qd}  and Unparticles~\cite{Martinez:2008hm}.
CMDM and CEDM are defined through the effective Lagrangian
 \begin{equation}
\Lagr= \bar{u}(t) \frac{-g_s}{2m_t} \sigma_{\mu\nu}G^{\mu\nu a } T^a \left(\Delta\tilde{\kappa}+i \gamma_5\Delta\tilde{d} \right)  u(t),
\label{lgr}
\end{equation}
where $\Delta\tilde{\kappa}$ and $\Delta\tilde{d}$ represent the CMDM and CEDM respectively, $G^{\mu\nu a}$ is the gluon field strength, and $T^a$ are the QCD fundamental generators of SU(3)$_c$.

In a recent study by the CMS collaboration spin correlation  in $t\bar{t}$ cross section is used to obtain a new bound to CMDM; the bound found is $\text{Re}({\Delta\tilde\kappa})=0.037\pm0.041$, at 95$\%$CL, or equivalently $-0.045<\text{Re}({\Delta\tilde\kappa})<0.119$~\cite{cmstop}. The given bound is obtained in the dilepton channel in $pp$ collisions at $\sqrt{s}=7$~TeV with an
integrated luminosity of $5$~fb$^{-1}$ and is compared with the SM theoretical prediction,  including  a
new physics contribution.
Bounds to the top quark CMDM and CEDM are found from combined results from Tevatron and LHC(Atlas)
using high values of the $m_{t\bar{t}}$ cross section, the bounds reported are $|\Delta\tilde\kappa|<0.05$ and $\Delta\tilde{d}<0.16$ at 95$\%$C.L.~\cite{Kamenik:2011dk}.
It is estimated that {in order} to find a $5\sigma$ upper bound of $\Delta\tilde{d}<0.05$ it is required an integrated luminosity of
$10$~fb$^{-1}$ at $\sqrt{s}=13$~TeV at LHC~\cite{Gupta:2009wu}.

From CLEO data in the radiative decay $b\to s \gamma$ a strong constraint of  $-0.03<|\Delta\tilde\kappa|<0.01$ is obtained~\cite{Martinez:2001qs}. The SM
prediction to the CMDM is $\Delta\tilde\kappa\sim 5.6\times10^{-2}$~\cite{Martinez:2007qf}.
The  top quark CEDM and CMDM induce new contributions to the lightest quarks through the renormalisation group equations (RGE), therefore,
the neutron dipole moment gives an indirect bound to CEDM $|\Delta\tilde{d}|<1.9\times10^{-3}$~\cite{Kamenik:2011dk}.

 In the present work we want to study the THDM Type-III contribution, without imposing a $Z_2$ discrete symmetry, in the quark top CMDM and CEDM. This kind of model explicitly violates
 the CP symmetry in the scalar potential, which generates the top quark CMDM at one loop level. Explicit CP  violation generates mixing among the neutral CP-odd and CP-even scalar
 fields. This mixing is very suppressed by the recent data obtained for $R_{\gamma\gamma}$ in LHC, for the Higgs physics~\cite{BabarJohnWalshforthe:2014uda}.
Using the allowed region $\alpha_1- \alpha_2$ from the neutral scalar sector we find the allowed values  for  the top quark CDMD and CEDM.

\section{Two Higgs doublet model with CP violation}
\label{sec-intro}

The simplest extension of the SM, with one extra scalar doublet is called the two Higgs doublet model (THDM); the model contains two doublet fields, $\Phi_1$ and $\Phi_2$; this kind of model has the advantage of being  capable to describe the phenomenon of CP violation~\cite{Branco:2011iw}.
When a discrete symmetry is imposed there are two kind of models, in the so called Type-I one doublet gives mass to all quarks and in the Type-II model one doublet gives
mass to the up quarks while the other one gives mass to the down quarks. In a theory without the restriction of a discrete symmetry, also called THDM Type III, the two doublets simultaneously give
mass to the up and down quarks, and the mass matrix depends on the Yukawa couplings which cannot be simultaneously diagonalized, allowing the presence of flavour changing at tree level~\cite{Ginzburg:2004vp}.

If we consider a general THDM the scalar potential can be written as
\begin{eqnarray}
V &=&-\mu_{1}^{2}\Phi _{1}^{+}\Phi _{1}-\mu_{2}^{2}\Phi _{2}^{+}\Phi _{2}-\left[
\mu_{12}^{2}\Phi _{1}^{+}\Phi _{2}+h.c.\right]   \nonumber \\
&+&\frac{1}{2}\lambda _{1}\left( \Phi _{1}^{+}\Phi _{1}\right) ^{2}+\frac{1}{2}\lambda _{2}\left( \Phi _{2}^{+}\Phi _{2}\right) ^{2}
+\lambda _{3}\left( \Phi _{1}^{+}\Phi _{1}\right) \left( \Phi _{2}^{+}\Phi
_{2}\right) +\lambda _{4}\left( \Phi _{1}^{+}\Phi _{2}\right) \left( \Phi
_{2}^{+}\Phi _{1}\right)   \nonumber \\
&+&\left[ \frac{1}{2}\lambda _{5}\left( \Phi _{1}^{+}\Phi _{2}\right)
^{2}+\lambda _{6}\left( \Phi _{1}^{+}\Phi _{1}\right) \left( \Phi
_{1}^{+}\Phi _{2}\right)  +\lambda _{7}\left( \Phi _{2}^{+}\Phi _{2}\right) \left( \Phi
_{1}^{+}\Phi _{2}\right) +h.c.\right] ,
\end{eqnarray}

where $\mu_{1}^2$, $\mu_{2}^2$, $\lambda_1$, $\lambda_2$, $\lambda_3$ and $\lambda_4$ are real parameters and the parameters $\mu_{12}^2$, $\lambda_5$, $\lambda_6$, and $\lambda_7$ can have complex values allowing the explicit CP violation in the potential.
The neutral components in the fields are defined as  $\frac{1}{\sqrt{2}}(v_a+\eta_a+i\chi_a)$, where $a=1,2$. The vacuum expectation values (VEV) can be taken real because complex phases
can be reabsorbed in the complex parameters in the scalar potential. The VEV take the values
\begin{equation}
\langle \Phi_1 \rangle= \frac{1}{\sqrt{2}}\left(
\begin{array}{c}
0 \\
v_1 \\
\end{array}
\right),
\label{vev1}
\end{equation}
\begin{equation}
\langle \Phi_2 \rangle= \frac{1}{\sqrt{2}}\left(
\begin{array}{c}
0 \\
v_2\\
\end{array}
\right).
\label{vev2}
\end{equation}

Due to the explicit CP symmetry breaking, there will be mixing among the CP-odd and CP-even scalar sectors.
Defining $\tan\beta=\frac{v_1}{v_2}$, we take the scalar field $(\eta_3=-\chi_1s_{\beta}+\chi_2c_{\beta})$
orthogonal to the Would-be Goldstone component corresponding to the  Z gauge boson.
After symmetry breaking, the mass eigenstates of the neutral Higgs bosons are related to the $\eta_{j}$ states as
\begin{equation}
h_{i}=\sum_{j=1}^{3}R_{ij}\eta _{j},
\label{h-Rn}
\end{equation}
where $i=1,2,3$ and the $R$ matrix is given by~\cite{Ginzburg:2004vp}:
\begin{equation}
R=\left(
\begin{array}{ccc}
c_{1}c_{2} & s_{1}c_{2} & s_{2} \\
-\left( c_{1}s_{2}s_{3}+s_{1}c_{3}\right)
& c_{1}c_{3}-s_{1}s_{2}s_{3} & c_{2}s_{3} \\
-c_{1}s_{2}c_{3}+s_{1}s_{3} & -\left(
c_{1}s_{3}+s_{1}s_{2}c_{3}\right)  &
c_{2}c _{3}
\end{array}
\right),
\end{equation}
where the abbreviations $c_i=\cos\alpha_i$ and $s_i=\sin\alpha_i$, with $i=1,2,3$ are used. $h_i$ eigenstates do not have a well defined CP state.
For convenience, we choose the neutral Higgs bosons $h_i$ to satisfy the mass hierarchy  $m_{h_1}\leq m_{h_2}\leq m_{h_3}$.
In the limit case when $s_2=s_3=0$ we recover the THDM without CP violation.

The Yukawa Lagrangian for the quark sector has the general form
\begin{eqnarray}
-\mathcal{L}_{Yukawa} &=&\sum_{i,j=1}^{3}\sum_{a=1}^{2}\left( \overline{q}%
_{Li}^{0}Y_{aij}^{0u}\widetilde{\Phi }_{a}u_{Rj}^{0}+\overline{q}%
_{Li}^{0}Y_{aij}^{0d}\Phi _{a}d_{Rj}^{0}  +h.c.\right ).
\label{yukawa}
\end{eqnarray}
In the above equation, $Y_{a}^{u,d,l}$ are the $3\times 3$ Yukawa matrices. $q_{L}$  denotes
the left handed quark doublets and $u_{R}$, $d_{R}$,  represent the right handed quark singlets under $SU(2)_{L}$.
After spontaneous symmetry breaking (SSB), the mass matrix can be written as
\begin{equation}
M^{u,d}=\sum_{a=1}^{2}\frac{v_{a}}{\sqrt{2}}Y_{a}^{u,d},  \label{mass}
\end{equation}
where $Y_a^{f}=V_L^f Y_a^{0f}\left(V_R^{f}\right)^\dag$, for $f=u,d$, and  $V_{L,R}^{f}$ are the rotation matrices that diagonalize the mass matrix.  The Yukawa matrix $Y_2^u$
as a function of $M^u$ and $Y^a_1$ gives THDM-II Lagrangian, with tree level flavor changing. For the up sector the Yukawa Lagrangian can be written as
\begin{eqnarray}
-\mathcal{L}_{Y}&=& \frac{1}{v\sin\beta} \sum_{ijk} \bar{u}_i M^u_{ij} \left(  A_k^u P_L + A_k^{*u} P_R \right)  u_j h_k  \nonumber \\
&+& \frac{1}{\sin\beta}\sum_{ijk}\bar{u}_iY_{ij}^u  \left(  B_k^u P_L + B_k^{*u} P_R \right) u_j  h_k ,\nonumber \\
\label{yukawa_quarks}
\end{eqnarray}
where
\begin{eqnarray}
A_k^u&=& R_{k2} - i R_{k3}\cos\beta ,\nonumber \\
B_k^u&=& R_{k1} \sin\beta - R_{k2} \cos\beta + i R_{k3}.
\label{ak}
\end{eqnarray}
The $Y_{ij}$ also gives a contribution to the anomalous couplings CEDM and CMDM wich is of the same order of the one in the THDM-II coming from
$(\bar{u}_i M^u_{ij} \left(  A_k^u P_L + A_k^{*u} P_R \right)  u_j h_k )$.
If we asume the Cheng-Sher parametrization~\cite{Cheng:1987rs}, where $(Y_{tt}\simeq m_t/v)$, both contributions must be taken into account in order to compute $\Delta\tilde\kappa$ y $\Delta\tilde{d}$.
%
\begin{figure}[!htb]
\centering
\includegraphics[scale=.26]{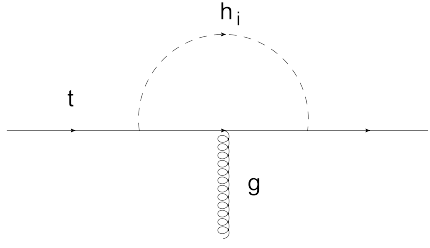}
\caption{Feynman Diagram for the anomalous quark-gluon couplings in the general THDM.}
\label{fig:00}
\end{figure}
\section{CMDM and CEDM in the general THDM}
\label{sec-2}
The anomalous couplings contributions for the CMDM $\Delta\tilde{\kappa}$ and for the CEDM delta $\Delta\tilde{d}$, arising from the
diagram in Fig.(\ref{fig:00}) are given by the following expressions
\begin{eqnarray}\label{eq:kt}
\Delta\tilde{\kappa}&=& \frac{G_{F}m_{t}^2}{2\sqrt{2}\pi^2 \sin^2{\beta}} \sum_{i=1}^{3} \int_{0}^{1} dx \int_{0}^{1-x}dy
\frac{1}{(x+y)^2-(x+y-1)\hat{m}_{h_i}^2} \times \\ \nonumber
&& \left[ (x+y)(x+y-1)(R_{i2}^2-\cos^2{\beta}R_{i3}^2)-(x+y)(R_{i2}^2+\cos^2{\beta}R_{i3}^2)\right],
\end{eqnarray}

\begin{eqnarray}\label{eq:dt}
\Delta\tilde{d}&=& - \frac{G_{F}m_{t}^2}{\sqrt{2}\pi^2 \sin^2{\beta}} \sum_{i=1}^{3} \int_{0}^{1} dx \int_{0}^{1-x}dy
\frac{(x+y)(x+y-1)}{(x+y)^2-(x+y-1)\hat{m}_{h_i}^2} \times \\ \nonumber & & \left[  \cos{\beta} R_{i3} R_{i2}\right] ,\end{eqnarray}
where $m_1$, $m_2$, and $m_3$ are the masses of the $h_1$, $h_2$, and $h_3$, respectively. In this calculations we have used the method presented in \cite{Martinez:1996cy,Martinez:2001qs,Martinez:2008hm}.

The contribution from the $Y_{ij}$ is
\begin{eqnarray}\label{eq:ktt}
\Delta\tilde{\kappa}_{tt}&=& \frac{G_{F}m_{t}^2}{2\sqrt{2}\pi^2 \sin^2{\beta}}  \sum_{i=1}^{3} \int_{0}^{1} dx \int_{0}^{1-x}dy
\frac{1}{(x+y)^2-(x+y-1)\hat{m}_{h_i}^2} \times \\ \nonumber
&&\left[(x+y)(x+y-1)\left( (R_{i1}\sin{\beta}-R_{i2}\cos\beta)^{2}- R_{i3}^2\right)
-(x+y) \left[ \left( R_{i1} \sin{\beta} - R_{i2}\cos{\beta}\right)^2 + R_{i3}^2\right]\right],
\end{eqnarray}
and for the CEDM  we have
\begin{eqnarray}\label{eq:dtt}
\Delta\tilde{d}_{tt}&=& - \frac{G_{F}m_{t}^2}{\sqrt{2}\pi^2 \sin^2{\beta}} \sum_{i=1}^{3} \int_{0}^{1} dx \int_{0}^{1-x}dy
\frac{(x+y)(x+y-1)}{(x+y)^2-(x+y-1)\hat{m}_{h_i}^2} \times \\ \nonumber  &(& R_{i1}\sin\beta-  R_{i2}\cos\beta )R_{i3}.
\end{eqnarray}
The contributions for CMDM and CEDM from the coupling proportional to $M^u$ in one vertex and $Y^{u}$ in the other vertex are given by
\begin{eqnarray}\label{eq:ktint}
\Delta\tilde{\kappa}_{int}&=& \frac{G_{F}m_{t}^2}{2\sqrt{2}\pi^2 \sin^2{\beta}}  \sum_{i=1}^{3} \int_{0}^{1} dx \int_{0}^{1-x}dy
\frac{1}{(x+y)^2-(x+y-1)\hat{m}_{h_i}^2} \times \\ \nonumber
&& \Big[ (x+y)(x+y-1)\left( (R_{i1}\sin{\beta}-R_{i2}\cos\beta)R_{i2} + R_{i3}^2\cos\beta \right)
\\ \nonumber &&	 - (x+y)\left( (R_{i1}\sin{\beta}-R_{i2}\cos\beta)R_{i2} - R_{i3}^2 \cos\beta \right) \Big],
\end{eqnarray}
and
\begin{eqnarray}\label{eq:dtint}
\Delta\tilde{d}_{int}&=& - \frac{G_{F}m_{t}^2}{\sqrt{2}\pi^2 \sin^2{\beta}} \sum_{i=1}^{3} \int_{0}^{1} dx \int_{0}^{1-x}dy
\frac{(x+y)(x+y-1)}{(x+y)^2-(x+y-1)\hat{m}_{h_i}^2} \times \\ \nonumber
&&\Big[ (x+y)(x+y-1)\left( R_{i2}R_{i3}- (R_{i1}\sin{\beta}-R_{i2}\cos\beta)R_{i3}\cos\beta \right)
\\ \nonumber &&	 - (x+y)\left( R_{i2}R_{i3} +  (R_{i1} \sin{\beta} - R_{i2}\cos{\beta})R_{i3}\cos\beta \right) \Big].
\end{eqnarray}
We are using the Cheng-Sher parametrization of $Y_{tt}=m_t/v$ in Eqs. (\ref{eq:kt}) - (\ref{eq:dtint}).
We denote three contributions as $\Delta\tilde{\kappa}_t = \Delta\tilde{\kappa}+\Delta\tilde{\kappa}_{tt}+ \Delta\tilde{\kappa}_{int}$
and $\Delta\tilde{d}_t = \Delta\tilde{d}+\Delta\tilde{d}_{tt} + \Delta\tilde{d}_{int} $.
The charged Higgs contribution to $\Delta\tilde{\kappa}$ and $\Delta\tilde{d}$ can be neglected
because in the loop circulates a bottom quark  and it is suppressed compared to the loop contribution in the neutral scalar sector where a top quark
is the circulating particle. For the second and third terms in the sum of Eqs. (\ref{eq:kt}) - (\ref{eq:dtint}) we considered $m_{h_2}=m_{h_3}=m_{H^+}$. In this partially degenerate case
there is CP violation for $\alpha_2\neq0$~\cite{ElKaffas:2006nt}.

We will study nine regions of interest in the $\alpha_1$-$\alpha_2$ parameter space, this approximate regions are described in Table \ref{tab:01} and are those already under consideration by the authors in previous work~\cite{Gaitan:2013yfa}. The allowed regions $R_{1..9}$ in the $\alpha_1-\alpha_2$ plane, are defined from experimental bounds
in $R_{\gamma\gamma}$~\cite{Basso:2012st}, where$R_{\gamma\gamma}$ is given by
\begin{equation}
R_{\gamma\gamma}=\frac{\sigma(gg\rightarrow h_1)Br(h_1\rightarrow\gamma\gamma)}{ \sigma(gg\rightarrow h_{SM})Br(h_{SM}\rightarrow\gamma\gamma)}.
\end{equation}
Because the charged Higgs contributes to the loop in $h_1\rightarrow\gamma\gamma$, the chosen values of $M_{H^\pm}$ and $\tan\beta$ affect the allowed region in $\alpha_1-\alpha_2$. The process $(B\to X_s \gamma)$ contains an important contribution from the charged Higgs, 
this process strongly restricts $M_{H^\pm}$ vs. $\tan\beta$~\cite{BabarJohnWalshforthe:2014uda}. For small values of $\tan\beta$
the bound to the charged Higgs mass is of around $300$~GeV~\cite{CMS:mxa}.  A global analysis of B decays restricts $M_{H^\pm}<400$~GeV and $\tan\beta<10$~\cite{Arbey:2012ax,Haisch:2012re,CMS:2012lza}.

In Table \ref{tab:01} are shown the $R_{i}$ regions for the given values of $M_{H^+}$
and ${\tan\beta}$. In each region we set the masses of the neutral Higgses $m_{h_2}$ and
$m_{h_3}$ equal to the mass of the charged Higgs $m_{h_2}=m_{h_3}=M_{H^+}$.

Using $(0.5\leq R_{\gamma\gamma} \leq2.0$), $M_{H^{\pm}}=300$~GeV and $\tan\beta=1.0$, the allowed regions in the $\alpha_1-\alpha_2$ plane are $R_1$ and $R_2$. For the same values in the other parameters and setting the charged Higgs mass to $M_{H^{\pm}}=500$~GeV region $R_3$ is obtained. Combining $(1\leq R_{\gamma\gamma}\leq2.0)$ with $M_{H^{\pm}}=350$~GeV
and  $\tan\beta=1.5$ the allowed region for  $\alpha_1-\alpha_2$ give $R_4$ and $R_5$. When $\tan\beta=2$ and for the same values to the other parameters as in $R_{4,5}$ we get
$R_{6,7}$. Finally if $\tan\beta=2.5$ the allowed regions are $R_{8,9}$.

\begin{figure}[!ht]
\centering
\includegraphics[scale=0.8]{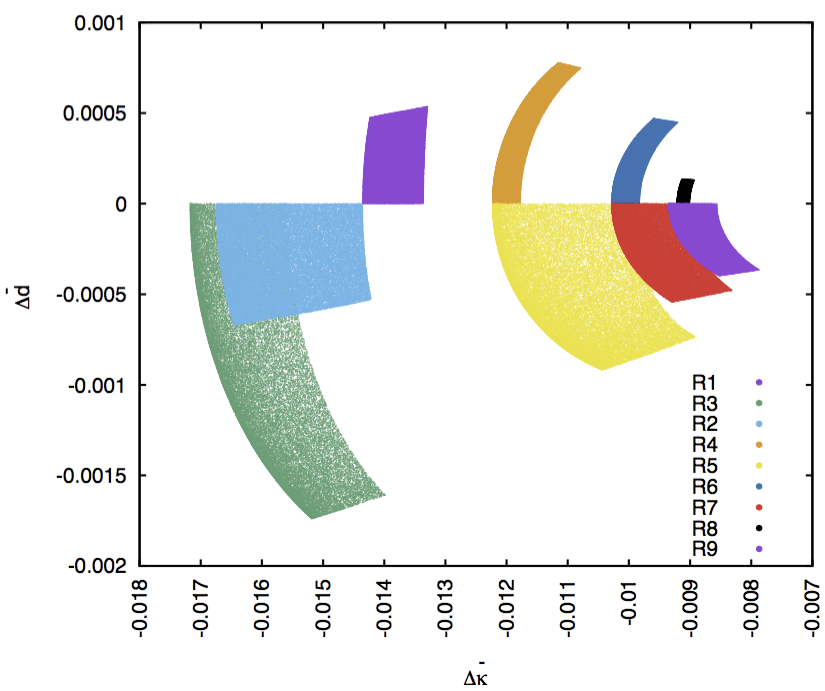}
\caption{Scatter plot in the  $\Delta\tilde{\kappa}$ and $\Delta\tilde{d}$ plane with random values of the angles $\alpha_1,\alpha_2$ in the range allowed for each region and $\alpha_3=0.$,  $\tan\beta$ and $M_{H^+}$ are as defined in Table (\ref{tab:01}) in every region.} \label{fig:03}
\end{figure}

\begin{figure}[htb]
\centering
\includegraphics[scale=0.8]{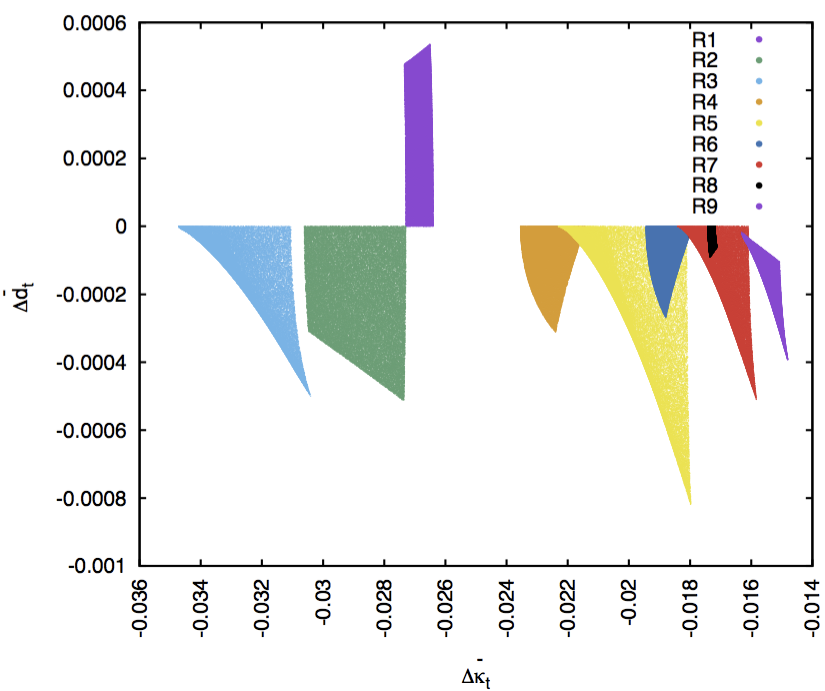}
\caption{Scatter plot in the  $\Delta\tilde{\kappa}_t$ and $\Delta\tilde{d}_t$ plane with random values of the angles $\alpha_1,\alpha_2$ in the range allowed for each region and $\alpha_3=0.$,  $\tan\beta$ and $M_{H^+}$ are as defined in Table (\ref{tab:01}) in every region. In this case we plot $\Delta\tilde{d}_t$ and $\Delta\tilde{\kappa}_t$ including the $Y_{tt}$ contribution.}\label{fig:04}
\end{figure}

\begin{figure}[!hb]
\centering
\includegraphics[scale=0.8]{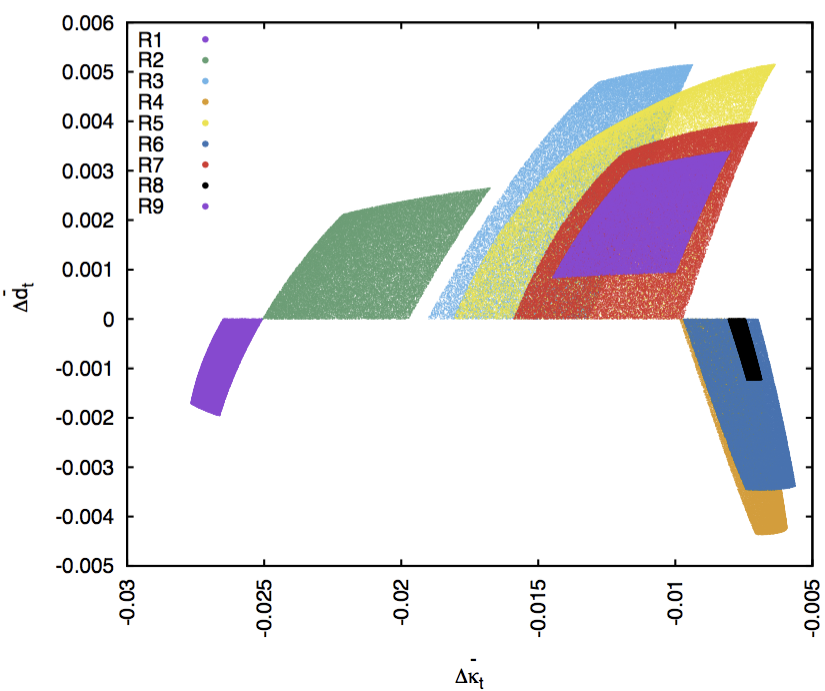}
\caption{Scatter plot in the  $\Delta\tilde{\kappa}_t$ and $\Delta\tilde{d}_t$ plane with random values of the angles $\alpha_1,\alpha_2$ in the range allowed for each region and $\alpha_3=0.$,  $\tan\beta$ and $M_{H^+}$ are as defined in Table (\ref{tab:01}) in every region. In this case we plot $\Delta\tilde{d}_t$ and $\Delta\tilde{\kappa}_t$ including the $Y_{tt}$ contribution and also the interference contribution.}\label{fig:05}
\end{figure}

In Table \ref{tab:02} we present the limits obtained for $\Delta\tilde{\kappa}_t$
and $\Delta\tilde{d}_t $ (or $\Delta\tilde{\kappa}$ and $\Delta\tilde{d}$) in a general THDM.
In order to illustrate the limits reported in Table \ref{tab:02} we show in Figs. (\ref{fig:03}), (\ref{fig:04}) and (\ref{fig:05}) the limits obtained for the anomalous moments, with values of the alpha parameters allowed for each region. In order to estimate the contribution of each term we separately analyze different cases. In the Fig. (\ref{fig:03}) only the contribution from Eqs. (\ref{eq:kt}) and (\ref{eq:dt}) is considered, meanwhile in Fig. (\ref{fig:04}) the contribution from $Y^{u}$, Eqs. (\ref{eq:ktt}) and (\ref{eq:dtt}), to previous values shown in Fig. (\ref{fig:03}) is added. All the contributions, including the interference terms (\ref{eq:ktint}) and (\ref{eq:dtint}), are considered in Fig. (\ref{fig:05}).
\begin{table}
\centering
\caption{$M_{H^{+}}$and $\tan \beta$ in each region. }
\begin{tabular*}{\columnwidth}{@{\extracolsep{\fill}}lllllll@{}}
\hline
 & $\alpha_{1}$  & $\alpha_{2}$ & $M_{H^{+}}$(GeV) & $\tan\beta$  \\
\hline
$R_1$ &  $0.67\leq \alpha _{1}\leq 0.8$  & $0\leq \alpha
_{2}\leq 0.23$ &  300& 1 \\

\hline
$R_2$ & $0.8\leq \alpha _{1}\leq 1.14$ & $-0.25\leq \alpha
_{2}\leq 0.$&  300&1  \\
\hline
$R_3$ & $1.18\leq \alpha _{1}\leq 1.55$& $-0.51\leq \alpha
_{2}\leq 0$ & 500 & 1 \\
\hline
$R_4$ & $ -1.57\leq \alpha _{1}\leq -1.3$& $ -0.46\leq
\alpha _{2}\leq 0.$ & 350 &1.5 \\
\hline
$R_5$ & $0.93\leq \alpha _{1}\leq 1.57$ &$  -0.61\leq
\alpha _{2}\leq 0.$  &350 &1.5 \\
\hline
$R_6$ & $-1.57\leq \alpha _{1}\leq -1.28$ & $ -0.38\leq
\alpha _{2}\leq 0.$ &350 & 2  \\
\hline
$R_7$ & $1.08\leq \alpha _{1}\leq 1.57$ & $-0.46\leq
\alpha _{2}\leq 0.$ &350 & 2  \\
\hline
$R_8$ & $ -1.39\leq \alpha _{1}\leq -1.3$ & $-0.13\leq
\alpha _{2}\leq 0.$ &350 & 2.5  \\
\hline
$R_9$ & $1.16\leq \alpha _{1}\leq 1.5$ &  $ -0.43\leq
\alpha _{2}\leq -0.1$&350 & 2.5  \\
\hline
\end{tabular*}
\label{tab:01}
\end{table}
\begin{table}
\centering
\caption{Range of values taken by the anomalous CMDM and CEDM. The first row shows only the contributions of $\Delta\tilde{\kappa}$ and $\Delta\tilde{d}$ according to Fig. (\ref{fig:03}). The regions in Fig. (\ref{fig:04}), which correspond to the addition of the  $Y^{u}$ contribution are shown in second row. The last row shows the regions for the three contributions based in Fig. (\ref{fig:05}).}
\begin{tabular*}{\columnwidth}{@{\extracolsep{\fill}}lllllll@{}}
\hline
R & CMDM &CEDM \\
\hline
$R_1$&  $-1.42\times10^{-2}  \leq \Delta\tilde{\kappa}  \leq -1.33\times10^{-2} $ & $0.  \leq \Delta\tilde{d}  \leq 2.41\times10^{-4}$\\
  & $  -2.73\times10^{-2}  \leq \Delta\tilde{\kappa}_t  \leq-2.63\times10^{-2}$ & $0.  \leq\Delta\tilde{d}_t  \leq  4.56\times10^{-4}$ \\
  &  $-2.76 \times10^{-2}  \leq \Delta\tilde{\kappa}_t  \leq -2.50 \times10^{-2} $ & $-1.26\times 10^{-3}   \leq \Delta\tilde{d}_t  \leq  0. $  \\
\hline
$R_2$ & $-1.67 \times10^{-2}  \leq \Delta\tilde{\kappa}  \leq -1.42\times10^{-2}$ & $-5.30\times10^{-4}  \leq \Delta\tilde{d}  \leq 0.$\\
 & $-3.06\times10^{-2}  \leq \Delta\tilde{\kappa}_t  \leq -2.37\times10^{-2}$&$-5.15\times10^{-4}   \leq\Delta\tilde{d}_t  \leq   0.$\\
& $-2.50 \times10^{-2}  \leq \Delta\tilde{\kappa}_t  \leq -1.67 \times10^{-2} $ &   $0.   \leq \Delta\tilde{d} _t \leq 2.65  \times10^{-3} $ \\
\hline
$R_3$ & $-1.72 \times10^{-2}  \leq \Delta\tilde{\kappa}  \leq -1.40 \times10^{-2} $ & $-1.61\times10^{-3}  \leq \Delta\tilde{d}  \leq 0.$\\
& $ -3.03\times10^{-2}   \leq \Delta\tilde{\kappa}_t  \leq  -2.43\times10^{-2}$ &$-6.63\times10^{-4} \leq\Delta\tilde{d}_t  \leq  0.$\\
&  $-1.35  \times10^{-2}  \leq \Delta\tilde{\kappa}_t  \leq -6.25 \times10^{-3} $ &  $0.   \leq \Delta\tilde{d}_t  \leq  5.13  \times10^{-3} $ \\
\hline
$R_4$ &$-1.18 \times10^{-2}  \leq \Delta\tilde{\kappa}  \leq -1.11 \times10^{-2}$  & $0.  \leq \Delta\tilde{d}  \leq 7.48 \times10^{-4}$ \\
& $-2.35\times10^{-2} \leq \Delta\tilde{\kappa}_t \leq -2.14\times10^{-2}$&$ -3.12\times10^{-4} \leq\Delta\tilde{d}_t  \leq  0.$\\
& $-9.84 \times10^{-3}  \leq \Delta\tilde{\kappa}_t  \leq -5.91  \times10^{-3} $ & $ -4.36\times10^{-3}  \leq \Delta\tilde{d}_t  \leq 0. $ \\
\hline
$R_5$ &$-8.89 \times10^{-3}  \leq \Delta\tilde{\kappa}  \leq -1.22 \times10^{-2} $ & $-7.36 \times10^{-4}  \leq \Delta\tilde{d}  \leq 0.$\\
&  $-1.79\times10^{-2}  \leq \Delta\tilde{\kappa}_t  \leq -2.23\times10^{-2}$&$-8.23\times10^{-4}\leq\Delta\tilde{d}_t  \leq 0.$ \\
& $-1.80 \times10^{-2}  \leq \Delta\tilde{\kappa}_t  \leq -6.33 \times10^{-3} $ &  $0.   \leq \Delta\tilde{d}_t  \leq  5.15  \times10^{-3} $ \\
\hline
$R_6$ & $-9.83 \times10^{-3}  \leq \Delta\tilde{\kappa}  \leq -9.59 \times10^{-3}$ &   $0 \leq \Delta\tilde{d} \leq 4.49 \times10^{-4}$  \\
 &$ -1.94\times10^{-2}  \leq \Delta\tilde{\kappa}_t  \leq  -1.79\times10^{-2}$&$-2.69\times10^{-4}\leq\Delta\tilde{d}_t  \leq 0. $\\
& $-9.71 \times10^{-3}  \leq \Delta\tilde{\kappa}_t  \leq -5.62 \times10^{-3} $ &  $-3.45  \times10^{-3}  \leq \Delta\tilde{d}_t  \leq  0. $\\
\hline
$R_7$ &$-1.03 \times10^{-2}  \leq \Delta\tilde{\kappa}  \leq -8.30 \times10^{-3}$  &$0 \leq \Delta\tilde{d} \leq -4.78 \times10^{-4}$ \\
& $-1.84\times10^{-2}  \leq \Delta\tilde{\kappa}_t  \leq  -1.58\times10^{-2}$&$-5.11\times10^{-4}\leq\Delta\tilde{d}_t  \leq 0.$\\
&  $-1.59  \times10^{-2}  \leq \Delta\tilde{\kappa}_t  \leq -6.99 \times10^{-3} $ &  $0.   \leq \Delta\tilde{d}_t  \leq 3.98  \times10^{-3}$ \\
\hline
$R_8$ & $-9.13 \times10^{-3}  \leq \Delta\tilde{\kappa}  \leq -9.01 \times10^{-3}$ &  $0. \leq \Delta\tilde{d} \leq 1.30 \times10^{-4}$ \\
&$  -1.74\times10^{-2} \leq \Delta\tilde{\kappa}_t  \leq -1.71\times10^{-2}$ & $-9.02\times10^{-5}\leq\Delta\tilde{d}_t  \leq 0.$\\
&  $-7.93 \times10^{-3}  \leq \Delta\tilde{\kappa}_t  \leq -6.55 \times10^{-3} $ &  $-1.23  \times10^{-3}  \leq \Delta\tilde{d}_t  \leq  0.$ \\
\hline
$R_9$ &$-9.30 \times10^{-3}  \leq \Delta\tilde{\kappa}  \leq -7.87 \times10^{-3}$  & $-3.65 \times10^{-4}  \leq \Delta\tilde{d}  \leq -1.04 \times10^{-4}$   \\
 & $-1.63\times10^{-2} \leq \Delta\tilde{\kappa}_t  \leq -1.48\times10^{-2}$ & $ -3.97\times10^{-4}\leq\Delta\tilde{d}_t  \leq -1.84\times10^{-5}$\\
&  $  -1.44   \times10^{-2}  \leq \Delta\tilde{\kappa}_t  \leq -8.0 \times10^{-3} $  & $8.31  \times10^{-4}  \leq \Delta\tilde{d}_t  \leq  3.59  \times10^{-3} $  \\
\hline
\end{tabular*}\label{tab:02}
\end{table}

\section{Conclusions}
In this work we have studied regions of interest in the $\alpha_1-\alpha_2$ parameter space, in order to calculate the contribution to
the top anomalous couplings CMDM and CEDM in the context of a general THDM with CP violation. In our analysis has been considered the
contribution of the Yukawa coupling $Y_{tt}$ using $M_{H^+}=300,350$~GeV and $\tan\beta=1,1.5,2,2.5$. We find for the nine regions of interest
that the value for $\Delta\tilde{\kappa}$  can be at most of order $10^{-2}$ and $\Delta\tilde{d}$ of order $10^{-4}$ . The contributions arising from the interference of $M^{u}$ and $Y_u$ have been considered in the results.
The contributions of $\Delta\tilde{\kappa}_t$ are added coherently and for the three contributions the variations are not appreciable for the different regions $R_i$, $i=1,...,9$. However the contribution of $\Delta\tilde{d}_t$ coming from the interference increases almost an order of magnitude for the regions $R_i$ and in some cases the sign is changed.

The recent NLO calculation for top-quark production including anomalous top-quark CMDM reports the bound $-0.0096<\Delta\tilde{\kappa}<0.0090$~\cite{Franzosi:2015osa}; our result for region $R_7$ is in agreement with this stringent constraint. We can also compare our theoretical bounds with those obtained from Higgs boson production at the LHC where more
conservative model independent bounds are obtained~\cite{Hayreter:2013kba}, our results in all nine regions are in agreement even with the most restrictive bounds projected at 14~TeV,  $-0.016<\Delta\tilde{\kappa}<0.008$ and $|\Delta\tilde{d}|<0.007$, as reported in Table \ref{tab:02} of reference~\cite{Hayreter:2013kba}.

With future LHC measurements at higher energy there will be an excellent chance to probe new physics properties of the top quark;
 anomalous dipole moments are a measure of this new physics properties that can also give some insight in the top quark structure.
A precise measurement of the top quark CMDM and CEDM, expected soon after future LHC results, will be a useful source of
information in order to discriminate among different SM extensions.
\acknowledgments {
This work is supported in part by PIAPI project in FES-Cuautitlan UNAM, Sistema
Nacional de Investigadores (SNI) in M\'exico. J.H. Montes de Oca is thankful for support from the postdoctoral CONACYT grant. R. M. thanks to COLCIENCIAS for the financial support.
E.A.G. thanks \textit{Programa de Becas Postdoctorales en la UNAM} and postdoctoral CONACYT grant. The authors thank L. Diaz-Cruz for useful comments to the manuscript.}

\end{document}